\documentclass[a4paper,11pt]{article}
\usepackage{pos}
\usepackage{amsmath}
\usepackage{amsfonts}
\usepackage{graphicx}
\usepackage{todonotes}
\usepackage{xspace}
\usepackage{hyperref}
\usepackage{siunitx}

\newcommand{\GeV}{\ensuremath{\,\text{GeV}}\xspace}

\newcommand{\sherpa}{S\protect\scalebox{0.8}{HERPA}\xspace}

\newcommand{\pythia}{P\protect\scalebox{0.8}{YTHIA}\xspace}

\newcommand{\amegic}{A\protect\scalebox{0.8}{MEGIC}\xspace}
\newcommand{\comix}{C\protect\scalebox{0.8}{OMIX}\xspace}
\newcommand{\CSS}{CSS\protect\scalebox{0.8}{HOWER}\xspace}

\newcommand{\rivet}{R\protect\scalebox{0.8}{IVET}\xspace}

\newcommand{\powheg}{P\protect\scalebox{0.8}{OWHEG}\xspace}

\newcommand{\LHAPDF}{L\protect\scalebox{0.8}{HA}P\protect\scalebox{0.8}{DF}\xspace}

\newcommand{\alphaS}{\alpha_\text{s}\xspace}

\newcommand{\MEPSatLO}{\text{\textsc{MEPS@LO}}\xspace}
\newcommand{\MEPSatNLO}{\text{\textsc{MEPS@NLO}}\xspace}

\newcommand{\MCatNLO}{\text{\textsc{MC@NLO}}\xspace}

\newcommand{\EIC}{EIC\xspace }
\newcommand{\HERA}{HERA\xspace}
\newcommand{\ZEUS}{ZEUS\xspace}

\renewcommand{\logo}{\relax}

\title{Event generation for future DIS experiments}

\author[a]{Peter Meinzinger}
\author*[b]{Daniel Reichelt}
\author[c]{Federico Silvetti}

\affiliation[a]{Physik-Institut, Universit\"at Z\"urich, Winterthurerstrasse 190, CH-8057 Z\"urich, Switzerland}
\affiliation[b]{CERN, Theoretical Physics Department, CH-1211 Geneva 23, Switzerland}
\affiliation[c]{Institute for Particle Physics Phenomenology, Department of Physics, Durham University, Durham DH1 3LE, United
Kingdom}

\emailAdd{d.reichelt@cern.ch}

\abstract{
  \begin{picture}(0,0)
    \put(420, 350){\makebox(0,0)[r]{\small\tt CERN-TH-2026-140}}
  \end{picture}%

  In this contribution we discuss state-of-the-art hadron-level predictions for the deep-inelastic scattering process at next-to-leading-order precision for several multiplicities, consistently merged in one sample. We focus on the physics at (potential) future colliders, the Electron-Ion Collider planned at BNL as well as the higher energy experiments discussed as future options at CERN, a LHeC and a DIS phase of the Future Circular Collider dubbed FCC-eh.}

\FullConference{
33rd International Workshop on Deep Inelastic Scattering (DIS2026)\\
4-8 May 2026\\
Bologna, Italy\\
}

\begin{document}
\maketitle

\section{Introduction}

Planning and operation of next-generation lepton-hadron colliders requires precise theoretical predictions for the associated final states.
Experiments such as those conducted at \HERA~\cite{Klein:2008di,Abramowicz:1998ii,Newman:2013ada} have demonstrated the power of this class of processes as probes of hadron structure.
In the DIS regime, rigorous factorization theorems~\cite{Ellis:1978ty, Collins:1989gx} allow the cross section to be expressed as a convolution of perturbatively calculable hard-scattering matrix elements and non-perturbative parton distribution functions (PDFs).
The physics opportunities offered by lepton-hadron collisions have attracted renewed attention due to several proposed next-generation facilities.
The most advanced of these projects is the Electron-Ion Collider (\EIC) at Brookhaven National Laboratory, which currently serves as a major driver for theoretical and phenomenological studies.
Its scientific program aims to characterization of the proton's internal structure at an unprecedented level of precision, including both the spatial distribution of partons and their transverse-momentum dynamics~\cite{Accardi:2012qut, AbdulKhalek:2021gbh}.

In parallel, initiatives for future experiments at CERN have explored the possibility of future lepton-hadron collision programs.
The proposed Large Hadron-Electron Collider (LHeC) would operate using the existing Large Hadron Collider infrastructure and achieve center-of-mass energies around $\sqrt{s} \approx 1.2~{\rm TeV}$.
Looking further ahead, the Future Circular Collider, could enable DIS studies at energies up to $\sqrt{s} \approx 3.5~{\rm TeV}$.
Extracting the maximum scientific benefit from these facilities requires a reliable description of the hadronic final state.
Such precision is essential both for the design of future measurements and for the interpretation of experimental data.

In modern Monte Carlo event generators, multijet configurations are commonly modeled using merging algorithms that combine matrix elements of several partonic multiplicities with parton-shower evolution.
This work employs the CKKW merging approach within the \sherpa~\cite{Sherpa:2024mfk} event generator.
Originally developed at leading order (LO)~\cite{Hoeche:2009rj,Carli:2010cg}, the method was later generalized to next-to-leading order (NLO) accuracy~\cite{Hoeche:2012yf,Gehrmann:2012yg}.
A related study of LO multijet merging for DIS has recently been carried out in the \pythia framework~\cite{Helenius:2024wjg}.
Although such merged simulations do not constitute a complete higher-order calculation, they are particularly effective in describing the dominant dynamics at low virtualities, especially for $Q^2 \simeq \mathcal{O}(1\text{-}10\GeV^2)$.

Phenomenological studies of DIS at the \EIC have, for the most part, relied on leading-order matrix elements matched to parton showers (LO+PS), with \pythia being the tool of choice in many analyses~\cite{Page:2019gbf,Chien:2021yol,Arratia:2019vju,Arratia:2020nxw,Zheng:2018ssm,Arratia:2020azl,Arratia:2022oxd}.
Especially Ref.~\cite{Page:2019gbf} examined the entire range of virtualities accessible at the \EIC at LO.
Progress beyond LO has recently been achieved through DIS simulations matched at NLO within the \powheg framework~\cite{Banfi:2023mhz,Borsa:2024rmh,Buonocore:2024pdv}.
In addition, several computational tools are now capable of producing predictions for \EIC observables at next-to-next-to-leading order (NNLO) accuracy~\cite{Karlberg:2024hnl,NNLOJET:2025rno}.
The photoproduction regime has also been studied at NLO for jet production processes, as reported in Refs.~\cite{Meinzinger:2023xuf, Andersen:2024czj}.

In this contribution, we summarize the results of Ref.~\cite{Meinzinger:2025pam}, which presents the first MEPS@NLO predictions for both neutral- and charged-current DIS at the \EIC, and extend the study to the higher-energy LHeC and FCC-eh scenarios.

\section{The \sherpa event generator framework}

The results presented in this work are obtained using the \sherpa~3 event generator~\cite{Sherpa:2024mfk}.
Tree-level matrix elements are provided by the internal generators \amegic~\cite{Krauss:2001iv} and \comix~\cite{Gleisberg:2008fv}.
The corresponding virtual contributions for DIS processes are also implemented intenally.
Parton-shower evolution is described using \sherpa's default \CSS~\cite{Schumann:2007mg}, which is based on the Catani--Seymour dipole formalism~\cite{Catani:1996vz}.
Matching between NLO matrix elements and the parton shower is performed using the \MCatNLO prescription~\cite{Frixione:2002ik}, and matrix elements with different final-state multiplicities are consistently combined through the CKKW merging procedure~\cite{Catani:2001cc}.

For the non-perturbative stage of the event simulation, hadronization is carried out either with the cluster fragmentation model~\cite{Webber:1983if} as implemented in \sherpa~\cite{Winter:2003tt,Chahal:2022rid} or, alternatively, through the Lund string model via the interface to \pythia~8~\cite{Bierlich:2022pfr}.
Event analysis is performed using the built-in interface to \rivet~\cite{Bierlich:2019rhm}.

Parton densities are taken from the NNPDF30\_nlo\_as\_0118 set~\cite{NNPDF:2014otw}, accessed through \LHAPDF~\cite{Buckley:2014ana}.
The strong coupling is evaluated consistently with the PDF, in particular yielding $\alphaS(M_Z)=0.118$.

The automation of DIS matching and merging in \sherpa was introduced in Ref.~\cite{Carli:2010cg}.
In the present study, higher-order matching relies on \sherpa's implementation of the \MCatNLO formalism~\cite{Hoeche:2011fd}.
We employ the multijet-merging approaches known as \MEPSatLO~\cite{Hoeche:2009rj} and \MEPSatNLO~\cite{Hoeche:2012yf,Gehrmann:2012yg}.
Detailed descriptions of these algorithms and the associated treatment of NLO matrix elements can be found in the original references.

For the \MEPSatNLO setup, matrix elements for single- and dijet production are included at NLO accuracy, while three- and four-jet final states are incorporated at LO.
In neutral-current DIS this corresponds to $e^-p\to e^- + 1,2,j\,@\,\mathrm{NLO}+3,4,j\,@\,\mathrm{LO}$ and, for charged-current interactions, to
$e^-p\to \nu +1,2,j\,@\,\mathrm{NLO}+3,4,j\,@\,\mathrm{LO}$. Analogous but reduced process sets are used for the \MEPSatLO, \MCatNLO, and pure LO predictions.

In the neutral-current simulation, the light-quark flavors are treated as massless.
Separate LO matrix elements containing massive charm and bottom quarks are additionally generated and merged.
In either case we use the five-flavor NNPDF30\_nlo\_as\_0118 PDF set.
Consistently, the parton shower employs massive splitting kernels for both $c$ and $b$ quarks~\cite{Catani:2002hc}.
For charged-current DIS, only four massless quark flavors are considered.
Contributions involving bottom quarks are neglected since a diagonal CKM matrix is assumed, and any residual effects are strongly suppressed by the PDFs.

The merging procedure requires each higher-multiplicity final state to be clustered back to an underlying $2\to2$ topology.
Each event is assigned to one of three possible core processes, which define the characteristic scale $\mu_{\mathrm{core}}$:
\begin{enumerate}
\item[(i)] DIS-like scattering through virtual-photon exchange,
$ej\to ej$, with
$\mu_{\mathrm{core}}^2=Q^2$
\item[(ii)] photon-parton scattering,
$\gamma^*j\to j_1j_2$, where
$\mu^2_{\mathrm{core}}
= m_{\perp,1}m_{\perp,2}$,
defined as the product of the transverse masses of the outgoing jets,
$m_{\perp,i}=\sqrt{m_i^2+p_{\perp,i}^2}$
\item[(iii)] purely QCD-induced scattering,
$jj\to jj$, for which
$\mu^2_{\mathrm{core}}
=-\frac{1}{\sqrt{2}}
\left(s^{-1}+t^{-1}+u^{-1}\right)^{-1}$,
corresponding to a scaled harmonic mean of the Mandelstam invariants
$s$, $t$, and $u$.
\end{enumerate}
The factorization, renormalization, and shower starting scales are all chosen equal to this core scale,
\begin{equation}
\mu_{\mathrm{F}} =  \mu_{\mathrm{R}} = \mu_{\mathrm{Q}} = \mu_{\mathrm{core}}.
\end{equation}
The separation between the matrix-element and parton-shower domains is controlled by the merging scale $Q_{\mathrm{cut}}$.
Rather than fixing this quantity globally, a dynamic definition is employed,
\begin{equation}
Q_{\mathrm{cut}} =
\frac{\bar{Q}_{\mathrm{cut}}}
{\sqrt{1+\frac{\bar{Q}^2_{\mathrm{cut}}}
{S_{\mathrm{DIS}}Q^2}}}
\label{eq:Qcut}
\end{equation}
with $\bar{Q}_{\mathrm{cut}}=25\GeV$ and $S_{\mathrm{DIS}}=0.4$. The parameter $\bar{Q}_{\mathrm{cut}}$ prevents the unnecessary generation of high-multiplicity matrix elements in regions where they are not required, while the $Q^2$ dependence introduced through $S_{\mathrm{DIS}}$ lowers the merging scale at small virtualities and thereby improves the description of this kinematic regime.

\section{DIS at HERA and the EIC}

Neutral-current DIS within the \sherpa framework has been explored in a number of previous studies, including Ref.~\cite{Carli:2010cg}.
More recently, early releases of \sherpa~3 have been used by the H1 Collaboration in several analyses of event-shape and jet observables~\cite{H1:2023fzk, H1:2024nde, H1:2024aze, H1:2024pvu, H1:2024mox}.
Across these measurements, the generator generally provided a satisfactory description of the experimental data.
The \MEPSatNLO configuration adopted in the present work is closely related to that used in Refs.~\cite{Knobbe:2023ehi,Knobbe:2024rci}, where predictions were benchmarked against analytic resummation calculations for DIS event-shape observables.
Those studies also determine the non-perturbative parameters related to beam fragmentation that are incorporated in the current tuning.
Taken together, these provide a strong validation of the \sherpa framework and the \MEPSatNLO merging approach especially in NC DIS.
In \cite{Meinzinger:2025pam} we have further validated the simulation of CC DIS with data from the \ZEUS experiment \cite{ZEUS:2008arl} and studied the impact of merging  at LO and NLO in the kinematic regime accessible at the \EIC.
We found large corrections, up to a factor of 2, in going from leading order or NLO matched setups to the merged \MEPSatLO and \MEPSatNLO schemes in the low $Q^2$ and $x$ phase space.
Virtual corrections included in the NLO samples led to observable effect predominantly at high $Q^2$ and $x$.

\section{DIS at the LHeC}

\begin{figure}[!ht]
    \centering
    \includegraphics[width=0.49\linewidth]{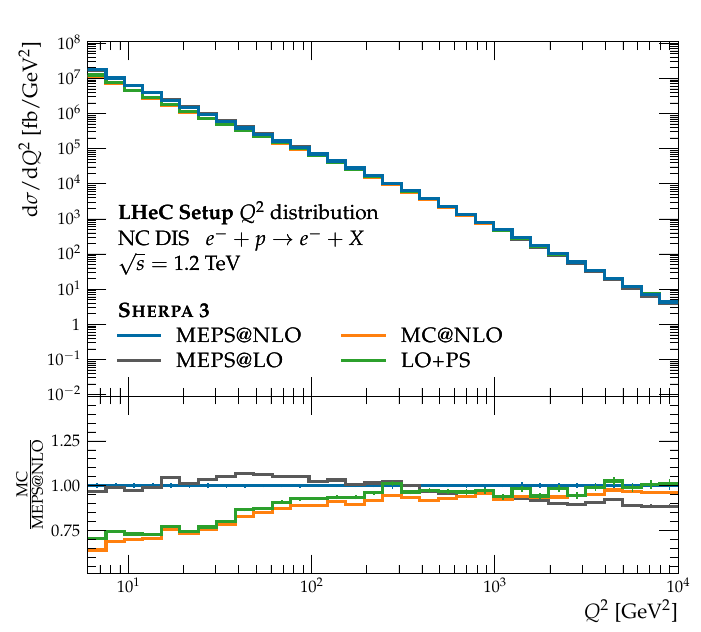}
    \includegraphics[width=0.49\linewidth]{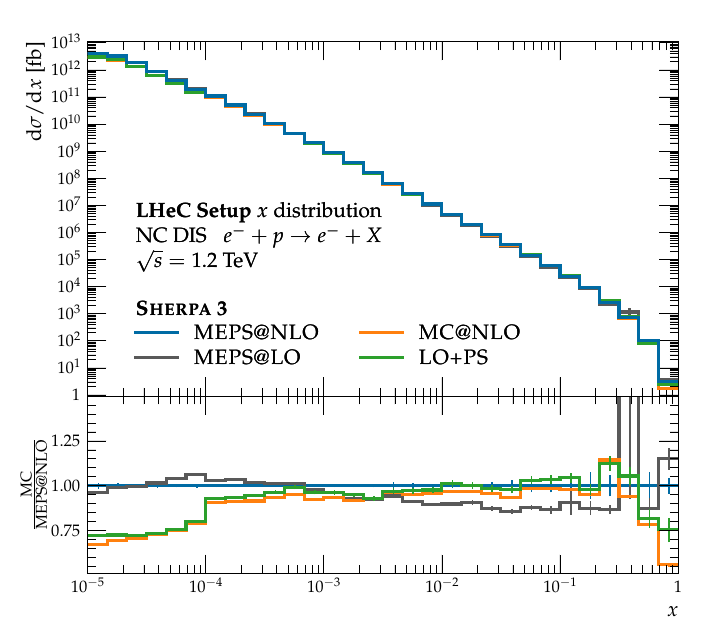}
    \caption{Distribution of $Q^2$ (left) and $x$ (right) for an LHeC setup at various MC accuracies.}
    \label{fig:LHeC-Q2-x}
\end{figure}

\begin{figure}[!h]
    \centering
    \includegraphics[width=0.49\linewidth]{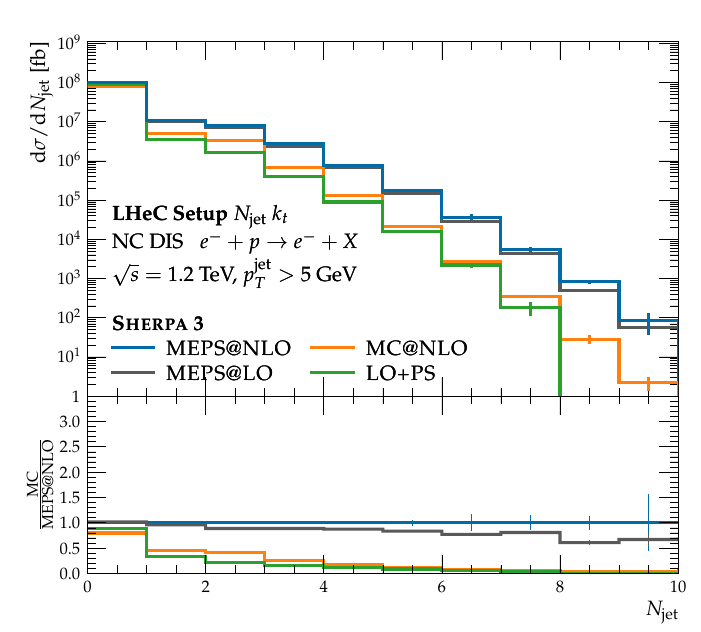}
    \includegraphics[width=0.49\linewidth]{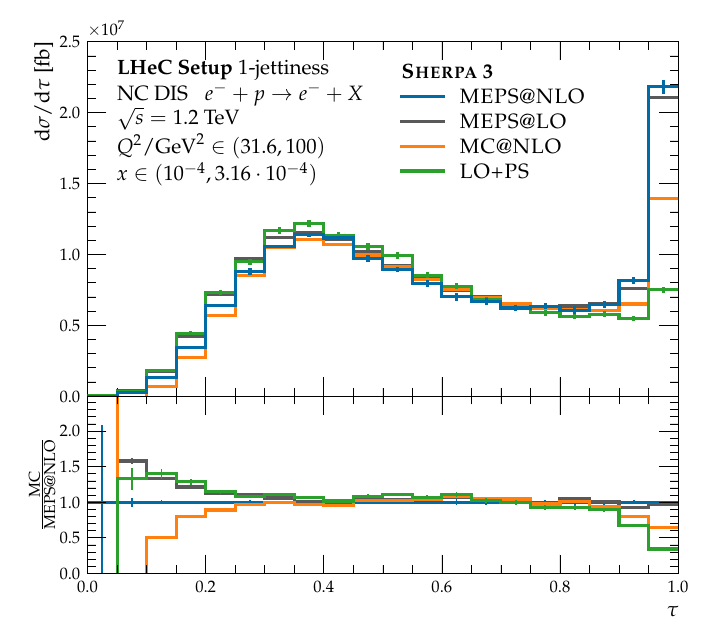}
    \caption{Distribution of jet multiplicity (left) and 1-jettiness $\tau$ (right) for an LHeC setup at various MC accuracies.}
    \label{fig:LHeC-N-tau}
\end{figure}

In Fig.~\ref{fig:LHeC-Q2-x} we show the distribution of the momentum transfer $Q^2$ and Bjorken-$x$, obtained from Monte Carlo samples of various accuracies.
We observe that the MC@NLO corrections to the leading order distribution are rather small, however the LO merged sample shows the characteristic corrections at small $Q^2$ and small $x$ associated with the appearance of additional scales like jet transverse momenta, dominating over the rather small $Q^2$.
The additional corrections in the NLO merged sample, \MEPSatNLO, are again rather small compared to the correction from merging.
This is in line with the effects observed for the EIC, however now stretching to significantly higher $Q^2$ (as might be expected from the increased center of mass energy).
For this setup we observe significant effects up to $100$-$200~\GeV^2$.
Only from $\approx 1000~\GeV^2$ does the \MEPSatNLO agree better with the \MCatNLO than with the \MEPSatLO, a sign that merging effects become less important and instead virtual corrections becoming significant.
In $x$ we observe a similar picture.
Additionally, we consider the jet multiplicity and 1-jettiness observable in Fig.~\ref{fig:LHeC-N-tau}.
The jet multiplicity is defined as number of jets with a transverse momentum $p_T > 5~\GeV$ according to the $k_t$-algorithm.
As expected, the merged samples predict large corrections especially at high jet multiplicities, where the cross section predicted by the LO+PS or \MCatNLO samples is negligable compared to the merged cross section.
For the 1-jettiness we again observe most corrections in the small $\tau$ region, at least for the selected $Q^2$ and $x$ range, where the expected corrections to the overall cross sections are not at their extremes, as seen above. 

\section{DIS at the FCC-eh}
\begin{figure}[!ht]
    \centering
    \includegraphics[width=0.49\linewidth]{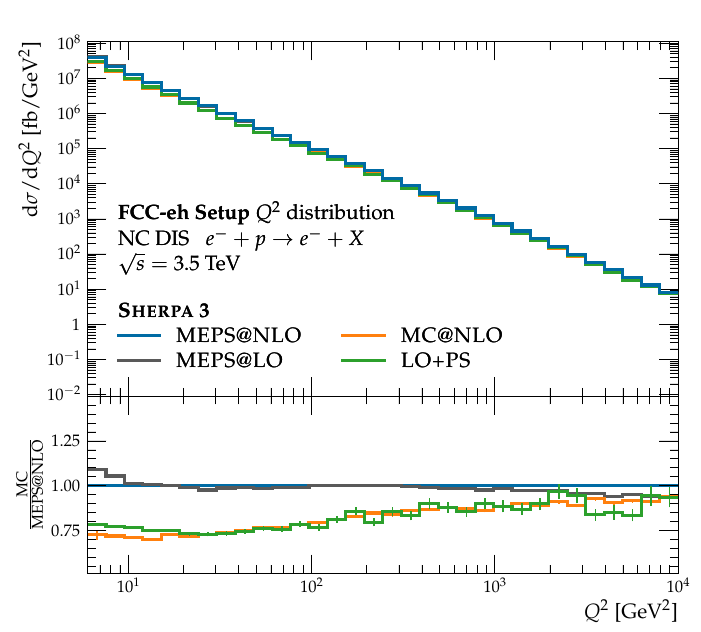}
    \includegraphics[width=0.49\linewidth]{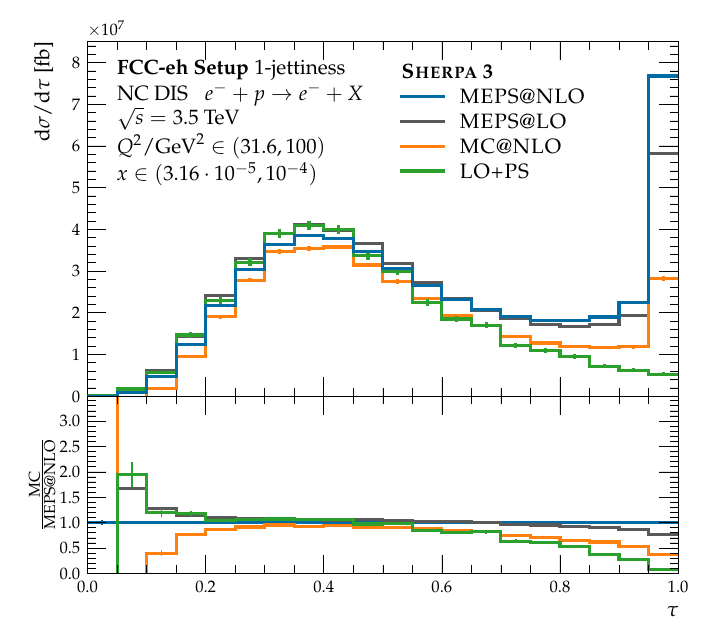}
    \caption{Distribution of $Q^2$ (left) and 1-jettiness $\tau$ (right) for an FCC-eh setup at various MC accuracies.}
    \label{fig:FCC-Q2-tau}
\end{figure}

We finally show predictions for the more speculative FCC-eh setup, at the highest center of mass energy.
We show the $Q^2$ and the 1-jettiness distribution in Fig.~\ref{fig:FCC-Q2-tau}.
Qualitatively we observe the same effect on the cross-section as a function of $Q^2$, but the large merging corrections now stretch to even higher $Q^2$ values.
Only at the edge of the plot range around $10^4~\GeV^2$ the predictions seem to be agreeing.
Of course, this comparison does not reflect the much larger range in $Q^2$ that would be accessible at such a machine.
Conversely, the 1-jettiness picture remains comparable to the LHeC setup for $\tau \leq 0.6$.
Instead, at higher $\tau$, merged samples show slightly stronger enhancement due to larger radiation phase space.

\section{Conclusion}

In this contribution, we have presented MEPS@NLO predictions for DIS at the LHeC and FCC-eh, complementing the detailed EIC study of Ref.~\cite{Meinzinger:2025pam}.
Across all collider scenarios, multijet merging induces significant corrections to the cross section at low virtualities.
This effect is driven by the appearance of additional hard scales, with these effects persisting to progressively higher $Q^2$ as the center-of-mass energy increases.
Merging corrections are a crucial ingredient for phenomenology at future DIS experiments, especially in the low $Q^2$ region and for the interpolation to the photoproduction regime.
As next steps, it would be desirable to have a fully consistent combination of photoproduction and DIS regimes, advanced treatment of QED corrections, and non-perturbative tunes utilising the full amount of available DIS data.

\section*{Acknowledgments}
D.R.\ is supported by the European Union under the HORIZON program in Marie
Sk{\l}odowska-Curie Project No. 101153541.
F.S.\ is supported by the STFC under Grant No. ST/P006744/1.
P.M.\ is supported by the Swiss National Science Foundation (SNF) under Contract
No. 200020-204200.

\bibliographystyle{JHEP}
\bibliography{journal,extra}

\end{document}